\documentclass{jpsj3}
\usepackage{txfonts}

\title{Nonadiabatic Fluctuations and the Charge-Density-Wave Transition in One-Dimensional Electron-Phonon Systems: a Dynamic Self-Consistent Theory}

\author{Alain~Mo\"ise~Dikand\'e$^1$\thanks{E-mail: dikande.alain@ubuea.cm}, and Claude~Bourbonnais$^2$}
\inst{$^1$Laboratory of Research on Advanced Materials and Nonlinear Sciences (LaRAMaNS), Department of Physics, Faculty of Science, University of Buea PO Box 63 Buea, Cameroon \\
$^2$Regroupement Qu\'eb\'ecois sur les Mat\'eriaux de Pointe, D\'epartement de Physique, Universit\'e de Sherbrooke, Sherbrooke Qu\'ebec Canada J1K-2R1} 

\abst{The Peierls instability in one-dimensional electron-phonon systems is known to be qualitatively well described by the Mean-Field theory, however the related self-consistent problem so far has only been able to predict a partial suppression of the transition even with proper account of classical lattice fluctuations. Here the Hartree-Fock approximation scheme is extended to the full quantum regime, by mapping the momentum-frequency spectrum of order-parameter fluctuations onto a continuous two-parameter space. For the one-dimensional half-filled Su-Schrieffer-Heeger model the ratio $d=\Omega/2\pi T_c^0$, where $\Omega$ is the characteristic phonon frequency and $2\pi T_c^0$ the lowest finite phonon Matsubara frequency at the mean-field critical point $T_c^0$, provides a natural measure of the adiabaticity of lattice fluctuations. By integrating out finite-frequency phonons, it is found that a variation of $d$ from the classical regime $d=0$ continuously connects $T_c^0$  to a zero-temperature charge-density-wave transition setting up at a finite crossover $d=d_c$. This finite crossover decreases within the range $0\leq d \approx 1$ as the electron-phonon coupling strength increases but remaining small enough for weak-coupling considerations to still hold. Implications of $T_c$ supression on the Ginzburg criterion is discussed, and evidence is given of a possible coherent description of the charge-density-wave problem within the framework of a renormalized Mean-Field theory encompassing several aspects of the transition including its thermodynamics close to the quantum critical point.}

\kword{Su-Schrieffer-Heeger model, quantum phonons, charge-density-wave transition, Hartree-Fock approach, renormalized Mean-Field theory, Ginzburg criterion}

\begin{document}
\maketitle

\section{Introduction}
\label{1}
Quasi-one-dimensional ($Q1D$) metals~\cite{toomb,jerome,gruner,monceau} become unstable against the electron-phonon interaction at low enough temperatures, leading to the formation of a charge density wave (CDW) and the opening of a gap at the Fermi level. The underlying mechanism~\cite{peierls,frohlich} involves a condensation of electron-hole pairs with total momentum $\pm 2\,k_F$ ($k_F$ is the Fermi wavector), due to the coupling of electronic states close to the Fermi surface to the $q=2\,k_F$ periodic lattice distorsion. This mechanism can be described in mean-field (MF)~\cite{toomb,jerome,gruner,monceau,rice,schulz1} using $1D$ tight-binding models such as the Su-Schrieffer-Heeger (SSH) model~\cite{suh} with dispersive phonons or the molecular-crystal model~\cite{holstein,dikad}. With these models the MF theory predicts a structural instability at a critical temperature $T_c^0$ manifest through softening of $2k_F$ phonons and the appearance of a gap driven all together by phonon-mediated electronic correlations. \\
For systems with weak electron-phonon interactions~\cite{eliash,mcmillan} the MF picture gives satisfactory account of key qualitive features of the CDW instability, in particular the possible mapping of the related self-consistent problem onto a classical field-theory problem involving a Ginzburg-Landau (GL) free energy~\cite{toomb,jerome,rice,schulz1,mckenzie,dietrich} has been quite appealing for the broad range of characteristic features of the transition which can be explored, extending from the thermodynamics~\cite{schulz1,dietrich} to signatures of the transition in single-particle properties~\cite{rice,schulz1}. However in real $Q1D$ materials the CDW instability is actually preceded by important transient phenomena which cannot be adequately accounted for with the standard MF theory. Also quantum fluctuations, related to the dynamics of phonons, enhance zero-point motions of the lattice and therefore should promote near-zero critical phenomena in the $Q1D$ systems. \\ 
Several attempts to improve the classical MF picture in order to account for quantum fluctuations on the CDW transition have been made~\cite{zheng1,zheng2,gupta1,gupta2}, following pioneer Monte-Carlo simulations~\cite{hirsch1} as well as a semi-analytical description based on the two-cutoff renormalization-group~\cite{bourbon1} both in the zero-temperature regime. Most recently, ellaborate approaches that combine analytical and numerical treatments of the nonadiabaticity of phonons have been suggested including the Dynamic-Mean-Field (DMF) theory~\cite{blawid1}, which rests on a mapping of the composite electron-phonon lattice onto a single-site impurity problem, and an improved renormalization-group approach~\cite{tam,bakrim} that takes into account the non-instantaneous character of phonon-mediated electron-electron interactions. The dominant picture emerging from these studies is the strong suppression of the transition, with eventually a crossover~\cite{hirsch1,bourbon1,blawid1,tam,bakrim} to a quantum CDW phase in systems where phonon characteristic energies are high enough to compete with electronic bandwidths. \\
But besides finite-mass effects finite-temperature phenomena also enter into play in the CDW transition. In $Q1D$ materials typically both thermal and quantum fluctuations are present, and in some physical contexts may occur at competing energy scales. It should be recalled, concerning this last point, that not all CDW materials have a large characteristic phonon energy. Indeed in some of them (as for instance organic conductors and in general materials with wide electronic bandwidths) the energy scales involved in molecular vibrations are actually far below typical Fermi energies~\cite{jerome}. In such materials thermal fluctuations can compete very strongly with the lattice zero-point motion, and it is rather from this competition we must expect a sizeable suppression of finite-temperature critical phenomena in strict $1D$~\cite{mermin}. \\
On the theoretical side, finite (nonzero) values of the characteristic phonon frequency have emerged from past studies as the hallmark of nonadiabatic effects in low dimensional electron-phonon systems. However their formulation for the CDW transition has been carried out following different approaches, namely assuming different considerations on the lattice dynamics. Thus, in early works~\cite{hirsch1,bourbon1} the phonon dynamics is reduced to the zero-point motion neglecting contribution from phonon degrees of freedom in the intermediate frequency range. On the other hand the recent attempts rest essentially on the assumption of momentum-independent spectral functions, laying emphasis on their frequency dependences in sums over phonon degrees of freedom~\cite{blawid1,tam,bakrim}. If such assumption sounds reasonable for the Holstein model as the only nonzero frequency modes are optical phonons, in the SSH model context this represents a drastic approximation since phonons in this case are inherently dispersive. \\ 
The issue of a possible treatment of finite-frequency effects and momentum fluctuations of phonons within a unified framework has actually not been definitely addressed for the CDW instability. In fact the problem is quite complex, indeed considerations~\cite{kotliar} underlying the DMF theory confines its validity to infinite dimensional systems while the renormalization-group is yet to provide a coherent formulation of flow equations for relevant physical quantities such as the retarded electron-electron interactions, taking into consideration the simultaneous frequency and momentum dependences of phonon-mediated two-particle correlations in the different interfering electronic correlation channels~\cite{bakrim}. \\
In the present work we revisit the Hartree-Fock approximation scheme for Gaussian fluctuations on the CDW instability for the SSH model, our objective being to extend the ability of this approach through a systematic treatment of both frequency and momentum-dependent fluctuations. In this purpose we follow a combined analytical and numerical treatment of Gaussian fluctuations which keeps the frequency and momentum dependences of microscopic parameters in the Landau-Ginzburg-Wilson (LGW) functional, but also relax the constraint of their restriction at $T_c^0$ the benefit of which is well established~\cite{toomb,rice}. \\
Our scaling procedure in the separation of high-energy degrees of freedom from low-energy ones will put into play a two-parameter space~\cite{hertz}, which makes the sums over quantum degrees of freedom quite effective resulting in an adiabaticity parameter defined as the ratio of the characteristic phonon frequency $\Omega$ to the lowest finite phonon Matsubara frequency at the MF critical point (i.e. $2\pi T_c^0$). \\
In section~\ref{2} we introduce the model and derive a time-dependent LGW functional~\cite{hertz}, expressed in terms of relevant CDW auxiliary fields. Next we briefly review fundamental results of the classical MF theory, with particular emphasis on the MF critical temperature and the Ginzburg criterion. In section~\ref{3} we construct the dynamic Hartree-Fock approximation scheme for Gaussian fluctuations, and derive a dynamic self-consistent problem in which high-energy fluctuations are summed out in an appropriately defined two-parameter space. In section~\ref{4} we analyze results of numerical solutions to the dynamical self-consistent problem laying emphasis on the effects of variation of the adiabaticity parameter $\Omega/2\pi T_c^0$ on the transition temperature. Implications of $T_c$ correction on the system thermodynamics close to the renormalized critical temperature are discussed, in particular we examine the variation of the Ginzburg criterion as well as the specific heat jump for finite values of the adiabaticity parameter in the weak-coupling regime. Section~\ref{5} will be devoted to a summary of results and concluding remarks. 

\section{Generalized LGW functional and classical GL theory}
\label{2}
\subsection{The partition function and Generalized LGW functional}
\label{2a}
\begin{figure}[tb]
\begin{center}
\includegraphics[width=5.0in]{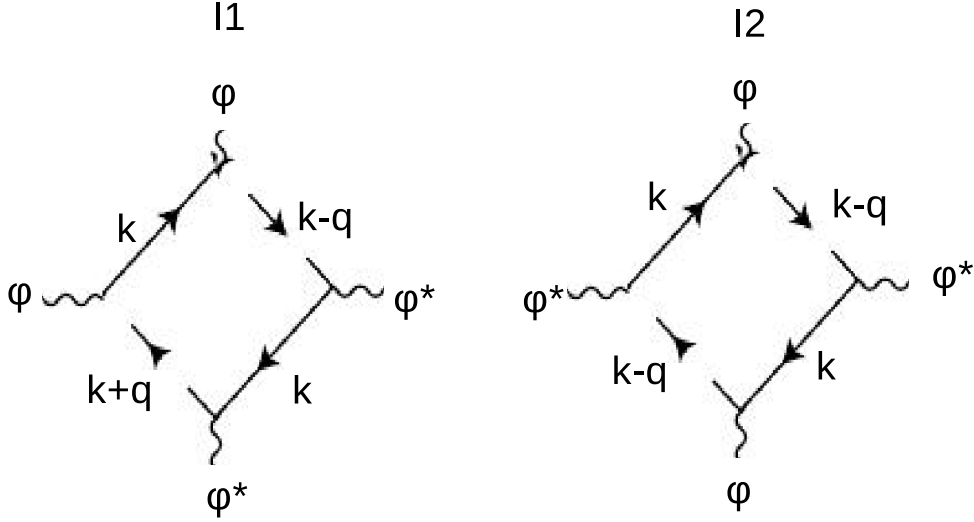}
\end{center}
\caption{The two distinct four-point vertices contributing to the dynamic mode-mode coupling matrix $\mathcal{B}(\tilde{Q}, T)$: dashed lines refer to left-moving electrons and solid lines to right-moving electrons.}
\label{figone}
\end{figure}

Consider the SSH model~\cite{suh} described by the Hamiltonian:
\begin{eqnarray}
H&=& H_0^e\, + H_0^{ph} + H_{e-ph}\nonumber \\
&=&\sum_{p,k,\sigma}{\epsilon_p(k)c_{p,k,\sigma}^{\dagger}c_{p,k,\sigma}} + \sum_{q}{\omega_q\left(b_q^{\dagger}b_q + 1/2\right)} \nonumber \\ 
&+&\frac{1}{\sqrt{L}}\sum_{\lbrace p,k,q,\sigma\rbrace }{g(k,q)c_{p,k+q,\sigma}^{\dagger} c_{-p,k,\sigma}\left(b_{-q}^{\dagger} + b_q\right)}, \nonumber \\
 \label{a1}
\end{eqnarray}
where $L$ is the length of the $1D$ electron-phonon system, $\epsilon_p(k)=\vartheta_F(pk-k_F)$ is the electronic dispersion linearized around the $1D$ Fermi points $pk_F=\pm k_F$, $\vartheta_F=2t\sin(ak_F)$ is the Fermi velocity and $\omega_q=\Omega\vert \sin(qa/2)\vert$ is the dispersion law for acoustic phonons with $\Omega=\omega_{\pi}$ the upper cut-off fixing the characteristic energy of lattice vibrations. \\
The first term of~(\ref{a1}) represents the free-electron Hamiltonian in which $c_{p,k,\sigma}^{\dagger}$ ($c_{p,k,\sigma}$) creates (annihilates) a $k$ electronic state with spin $\sigma$ in a $p$ electronic branch, the second term is the lattice contribution and the third term accounts for the interaction between a $k$ electron near the Fermi level and a $q$ phonon mode obeying the dispersion law $\omega_q$. In this last contribution, $g(k,q)$ is the electron-phonon interaction matrix defined as:
\begin{equation}
g(k,q)=\frac{4\alpha\,i}{\sqrt{2M\omega_q}}\sin(qa/2)\cos[(k+q/2)a], \label{a2}
\end{equation}
where $M$ is the mass of atoms/molecules, $\alpha$ is the electron-phonon coupling constant and $a$ is the lattice constant. \\ 
In momentum space and at half filling of the electronic band, the $q=\pi/a$ phonon mode coincides with the momentum exchange between pairs of perfectly nested electronic states on both side of the Fermi level. Consequently an electron in a $p$ branch can mix with a hole in the $-p$ branch leading to electron-hole pairs condensation in form of $2k_F$ CDW excitations driven by phonon-mediated two-particle correlations. \\
To formulate the $2k_F$ CDW excitation we start from the partition function $Z= Tr\,e^{-\beta H}$ (with $\beta= 1/T$, $k_B\equiv 1$), in which the trace over phonons involves harmonic degrees of freedom and hence is exact yielding the following effective partition function for the electrons in the functional-integral representation:   
\begin{equation}
Z_e=\int {[d\psi][d\psi^*]\,e^{S_e^0[\psi,\psi^*]}\,\exp{\left(-\int_0^{\beta}{d\tau_2\int_0^{\beta}{d\tau_1\,\mathcal{H}(\tau_2- \tau_1)}}\right)}}
\label{a3}
\end{equation}
where $\psi^{(*)}$ are Grassman fields,
\begin{equation}
S_e^0[\psi,\psi^*]=\sum_{p,k,\sigma}{\int_0^{\beta}{d\tau\psi_{p,\sigma}^*(k,\tau)G_p^0(k,\tau)\psi_{p,\sigma}(k,\tau)}} \label{a3a}
\end{equation}
is the free-electron action with 
\begin{equation}
G_p^0(k,\tau)=[-\partial/\partial \tau - \epsilon_p(k)]^{-1} \label{aa1}
\end{equation}
the free-electron propagator, and
\begin{equation}
\mathcal{H}(\tau_2- \tau_1)=-\int {dx\,g_{ph}(\tau_2- \tau_1)\mathcal{O}^*(x,\tau_1)\mathcal{O}(x,\tau_2)} \label{a4} 
\end{equation}
is an effective phonon-mediated electron-electron Hamiltonian. It is remarkable that we expressed explicitely this two-particle Hamiltonian in terms of density-wave fields $\mathcal{O}^{(*)}(x,\tau)$ which are defined as:
\begin{eqnarray}
\mathcal{O}^*(x,\tau)&=&\frac{1}{2}[O^*(x,\tau) - O(x,\tau)], \label{a5} \\
O(x,\tau)&=&\sum_{\sigma}{\psi_{-,\sigma}^*(x,\tau) \psi_{+,\sigma}(x,\tau)}, \nonumber
\end{eqnarray}
and which precisely, are composite fields associate with bond-order-wave (BOW) dynamics in the half-filled band electronic part of the  $1D$ electron-phonon system (\ref{a1})~\cite{dumoulin,nakamura}. Therefore, in terms of these fields, the two-particle Hamiltonian~(\ref{a4}) can readily be looked out as the governing Hamiltonian for BOW-BOW correlations in the $1D$ electron-phonon system. The phonon-mediated electron-electron coupling matrix associate with these BOW correlations is given by: 
\begin{eqnarray}
g_{ph}(\tau_2- \tau_1)&=& 2\vert g(k_F,2k_F)\vert^2\mathcal{D}_o(2k_F,\tau_2-\tau_1) \nonumber \\
                      &=& \frac{16\alpha^2}{M\Omega}\mathcal{D}_o(2k_F,\tau_2-\tau_1) \label{a6} 
\end{eqnarray}
and is dependent on the free-phonon propagator $\mathcal{D}_o(q,\tau_2-\tau_1)$, by virtue of the non-instantaneous (i.e. retarded) feature of the two-particle interaction. \\
We now consider correlations between BOW operators governed by the retarded electron-electron interaction Hamiltonian~(\ref{a4}), within the framework of the functional-integral formalism. In this purpose we introduce auxiliary quantum fields $\phi(x,\tau)$ conjugate to the BOW fields $\mathcal{O}(x,\tau)$. Applying a Hubbard-Stratonovich transformation~\cite{dumoulin,bourbon2} on the retarded two-particle Hamiltonian~(\ref{a4}), the effective electronic partition function~(\ref{a3}) becomes:  
\begin{equation}
Z_e=Z_0^e\,\int{[d\phi]\,\exp{\left(-\int{dx\int_0^{\beta}{d\tau_2\int_0^{\beta}{d\tau_1\,\phi(x,\tau_2)\vert g_{ph}(\tau_2-\tau_1)\vert^{-1}\phi(x,\tau_1)}}} +\,{\sum_{n=1}{\frac{1}{(2n)!}<S_{p}^{2n}>}}\right)}}, \label{a7}
\end{equation}
\begin{equation}
 S_{p}[\mathcal{O},\phi]=-2i\int{dx\int_0^{\beta}{d\tau\,\mathcal{O}(x,\tau)\phi(x,\tau)}}.  \label{a8}
\end{equation}
The cumulant series in~(\ref{a7}) involves only connected diagrams upon averaging over free electrons, to the fourth order in the auxiliary fields $\phi(x,\tau)$ this expansion gives $Z_e=Z_e^0\int{[d\phi]\,e^{-\beta \mathcal{F}[\phi]}}$ where:
\begin{eqnarray}
\beta\mathcal{F}[\phi]&=& \int{dx\int_0^{\beta}{d\tau_2\int_0^{\beta}{d\tau_1\,\phi(x,\tau_2)\left[\vert g_{ph}(\tau_2-\tau_1)\vert^{-1} + \chi(x, \tau_2-\tau_1)\right]\phi(x,\tau_1)}}} \nonumber \\ 
&+&\int{d\tilde{x}_1 \int{d\tilde{x}_2 \int{d\tilde{x}_3 \int{d\tilde{x}_4\, \mathcal{B}(\tilde{x}_1,\tilde{x}_2,\tilde{x}_3,\tilde{x}_4)\, \phi(\tilde{x}_1)\phi(\tilde{x}_2)\phi(\tilde{x}_3)\phi(\tilde{x}_4)}}}}, \label{a9}
\end{eqnarray}
is the LGW functional. Note that in~(\ref{a9}) we introduced the reduced variables $\tilde{x}\equiv (x,\tau)$ and the following quantities have been defined: 
\begin{equation}
\chi(x,\tau_2-\tau_1)= 2 <O(x,\tau_2)\,O(x,\tau_1)>, \label{a10a} 
\end{equation}
\begin{equation}
\mathcal{B}(\tilde{x}_1,\tilde{x}_2,\tilde{x}_3,\tilde{x}_4)= \frac{2}{3} <O(\tilde{x}_1)O(\tilde{x}_2)O(\tilde{x}_3)O(\tilde{x}_4)>, \label{a10b}
\end{equation}
corresponding respectively to two-point and four-point vertices associate with BOW-BOW correlation and mode-mode coupling matrix. \\
In Matsubara Fourier space the BOW-BOW correlation function $\chi(x,\tau_2-\tau_1)$ reads:
\begin{equation}
\chi(\tilde{Q},T)= 2 <O(\tilde{Q})\,O(-\tilde{Q})>, \label{a11}
\end{equation}
with $\tilde{Q}=(q-2k_F,\omega_m)$ where $\omega_m= 2m\pi T$ is phonon's Matsubara frequency. In fact this is nothing but the well-known $2k_F$ BOW susceptibility (see e.g. ref.~\cite{jerome}), which analytical expression after summing out the electronic variables $k$ and $\omega_n=(2n+1)\pi T$ turns to:
\begin{equation}
\chi(\tilde{Q}, T) = -N_F\lbrace \ln \frac{1.13 \, E_F}{T} + \Psi\left(\frac{1}{2}\right)- \frac{1}{2}\left[\Psi \left(\frac{1}{2} + \frac{\vert \omega_m\vert + i\vartheta_F Q}{4 \pi T}\right) + \Psi \left(\frac{1}{2} + \frac{\vert \omega_m\vert - i\vartheta_F Q}{4 \pi T}\right)\right] \rbrace, \label{a12}
\end{equation}
where $N_F=1/\pi\vartheta_F$ is the electron density of states at the Fermi level and $\Psi(...)$ is the Digamma function~\cite{hand1}. \\ 
As for the mode-mode coupling matrix, by Fourier transforming~(\ref{a10b}) and contracting the resulting normal-ordered product of fermion fields we are left with~\cite{leung,dikathesis}: 
\begin{equation}
\mathcal{B}(\tilde{Q}, T)= \frac{2}{3}\, \lbrack 2 I_1(\tilde{Q}, T) + I_2(\tilde{Q}, T)\rbrack \label{a13}
\end{equation} 
in which $I_1$ and $I_2$ are contributions from the first and second classes of four-point vertices respectively shown in figure~\ref{figone}. 
Their analytical expressions in terms of the free-fermion propagator $G_p^0(k,\omega_n)$ are straightforward i.e.:
\begin{eqnarray}
I_1(\tilde{Q}, T)&=& \left(\frac{T}{L}\right)^2 \sum_{\tilde{k}}{G_p^0(\tilde{k}-\tilde{Q}-2k_F)[G_{-p}^0(\tilde{k})]^2 G_p^0(\tilde{k}+\tilde{Q}+2k_F)}, \label{a14a} \\
I_2(\tilde{Q}, T)&=& \left(\frac{T}{L}\right)^2 \sum_{\tilde{k}}{[G_p^0(\tilde{k}+\tilde{Q}+2k_F)]^2[G_{-p}^0(\tilde{k})]^2}, \label{a14b}
\end{eqnarray}
or after summing out fermion variables ($k,\omega_n$):
\begin{eqnarray}
I_1(\tilde{Q}, T)&=& \frac{i\vartheta_F Q N_F}{16\pi L(\omega_m^2 + \vartheta_F^2Q^2)}\,\left[\Psi' \left(\frac{1}{2} + \frac{\omega_m + i\vartheta_F Q}{4 \pi T}\right) -\Psi' \left(\frac{1}{2} + \frac{ \omega_m - i\vartheta_F Q}{4 \pi T}\right)\right], \label{a15a} \\
I_2(\tilde{Q}, T) &=& \frac{TN_F}{2L(\omega_m^2 + \vartheta_F^2Q^2)}\,{\it Re}\left[\Psi \left(\frac{1}{2} + \frac{\omega_m + i\vartheta_F Q}{4 \pi T}\right) + \Psi \left(\frac{1}{2} + \frac{\omega_m - i\vartheta_F Q}{4 \pi T}\right)- 2\Psi\left(\frac{1}{2}\right)\right], \nonumber \\
\label{a15b} 
\end{eqnarray}
where $\Psi'(...)$ refers to the Trigamma function. 

\subsection{Classical limit of Ginzburg-Landau theory}
\label{2b}
In the classical regime where $Q=0$ and $\omega_m=0$, the BOW susceptibility~(\ref{a12}) is dominated by a logarithmic divergence at zero temperature. According to formula~(\ref{a15a}) and~(\ref{a15b}), in this regime $I_1(0, T)$ and $I_2(0, T)$ take the common value $-N_F\Psi''\left(\frac{1}{2}\right)/32\pi^2LT$. Also it should be stressed that in the classical MF picture BOW fields are static excitations, as a consequence we need to define appropriate (classical) auxiliary fields $\varphi$ from the (quantum) fields $\phi$. Thus we set $\phi=\varphi/\sqrt{T}$, and in terms of the classical fields now understood as BOW order parameters, the LGW functional $\mathcal{F}[\phi]$ in Matsubara Fourier space reduces to the GL functional:  
\begin{equation}
\mathcal{F}[\varphi]=\sum_Q {\left[\left(a + c_0^2 Q^2\right)\,\vert \varphi(Q)\vert^2 + \frac{b}{L}\,\vert \varphi(Q)\vert^4\right]}, \label{a16a}
\end{equation}
where microscopic parameters are defined as:
\begin{eqnarray}
a&=& N_F\,\ln\frac{T}{T_c^0}\approx a_0(\frac{T}{T_c^0} - 1), \hskip 0.2truecm a_0=N_F, \label{a16b} \\
c_0^2(T)&=& -\frac{N_F \vartheta_F^2}{2(4\pi T)^2}\,\psi''(1/2) \label{a16c} \\
&=&\frac{7\zeta(3)N_F \vartheta_F^2}{(4\pi T)^2}, \nonumber \\
b(T)&=& \left[c_0(T)/\vartheta_F\right]^2, \label{a16d}
\end{eqnarray} 
\begin{eqnarray}
T_c^0&=& 1.13\,E_F\,e^{-1/2\tilde{g}_0}, \label{a16e} \\
\tilde{g}_0&=&-\frac{8\alpha^2 N_F}{M\Omega}\mathcal{D}_o(2k_F, 0).
\end{eqnarray}
It turns out that in MF picture, the CDW instability is a second-order phase transition which occurs at a temperature  $T_c^o$. \\
If one ignores spatial fluctuations of the BOW order parameter, the homogeneous part of the GL functional~(\ref{a16a}) leads to two values of $\varphi$ minimizing the free energy. They are $\varphi=0$ for $T>T_c^0$ and $\varphi=\left[a_0/b(T_c^0)\right]^{1/2}\sqrt{T_c^0-T}$ for $T<T_c^o$, i.e. the system is disordered above $T_c^0$ and develops a bond-type CDW order below the critical temperature $T_c^0$. However, because of exisiting pretransitional processes fluctuations of the order parameter about its equilibra are actual relevant. Namely if we consider only the very small disturbances about the nonzero equilibrium, we find that while spatial fluctuations are thermally frozen they remain correlated with an exponentially decaying correlation function and a correlation length:   
\begin{equation}
\xi(T) = \xi_0 \sqrt{T_c^0}(T-T_c^0)^{-1/2}, \hskip 0.3truecm \xi_0= c_0(T_c^0)/(a_0)^{1/2}, \label{a17}
\end{equation}
diverging at $T=T_c^0$. Although this divergence appears to be inconsistent by carrying out a more exact analysis of the statistical mechanics of the second-order phase transition described by the GL theory via the transfer Matrix approach~\cite{scalap}, its existence however clearly reveals an important role of the characteristic temperature $T=T_c^0$ in the MF picture. In this context $T_c^0$ emerges from numerical transfer-matrix calculations~\cite{rice,scalap} as the characteristic temperature scale determining the validity of the classical MF theory, the associate criterion
\begin{equation}
\Delta t_0 = \left(\frac{2b(T_c^0)T_c^0}{a_0^2\xi_0}\right)^{2/3}, \label{a17a}
\end{equation}
i.e. the Ginzburg criterion (with $t=T/T_c^0$ the reduced temperature), defines the radius of the region around $T_c^0$ within which the correlation length is expected to diverge. It is generally assumed that the large $\Delta t_0$ the valid MF predeictions~\cite{jerome,schulz1}. Still, the Ginzburg criterion $\Delta t_0$ actually depends on microscopic paramaters and so its values cannot be arbitrarily chosen. In particluar for systems with a pronounced $1D$ structural anisotropy, values of $\Delta t_0$ which are fixed by the classical microscopic coefficients $a_0$, $b(T_c^0)$ and $c_0 (T_c^0)$ in~(\ref{a16b})-(\ref{a16d}) can hardly stabilize a true long-range CDW order in strict $1D$.

\section{The dynamic self-consistent theory}
\label{3}
\begin{figure}[tb]
\begin{center} 
\includegraphics{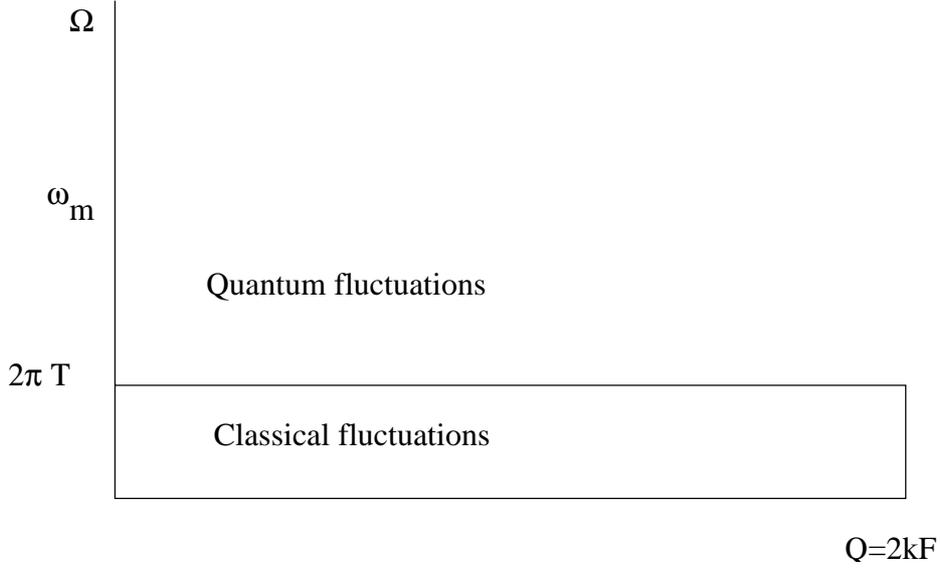}
\end{center}
\caption{Momentum-frequency map showing the high and low-energy regions in the Hartree-Fock approach to quantum fluctuations for the LGW functional~(\ref{a18}).}
\label{figtwo}
\end{figure}

\begin{figure}[tb]
\begin{center}
\includegraphics[width=5.0in]{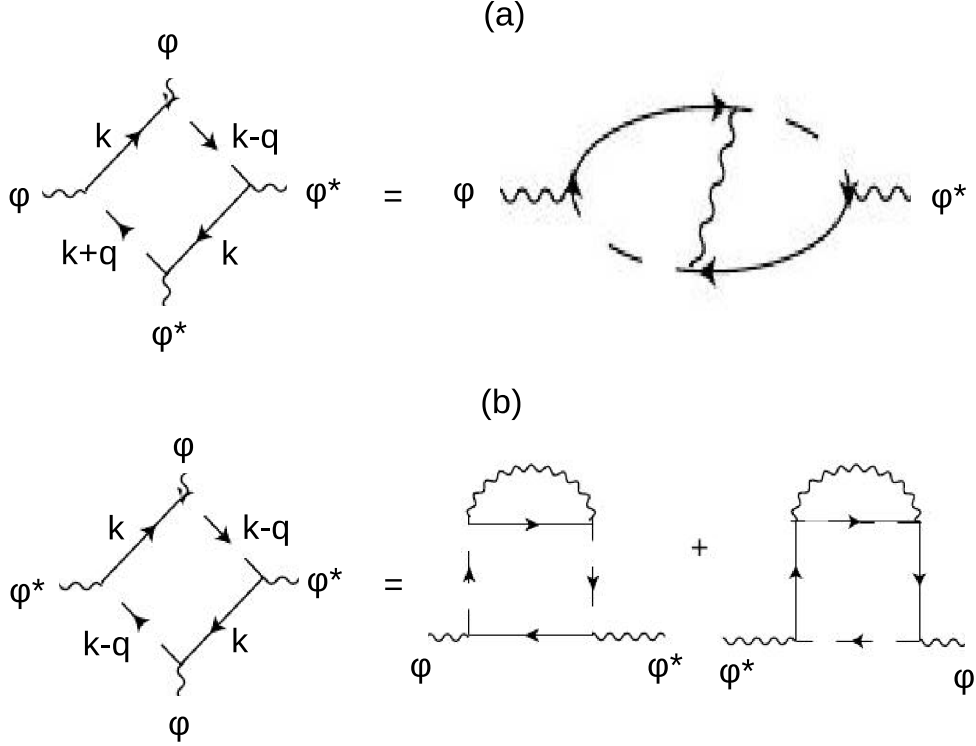}
\end{center}
\caption{Contributions to the self-energy from Gaussian contractions of the mode-mode coupling matrix $\mathcal{B}(\tilde{Q}, T)$.}
\label{figtwoa}
\end{figure}

\begin{figure}[tb]
\begin{center}
\includegraphics{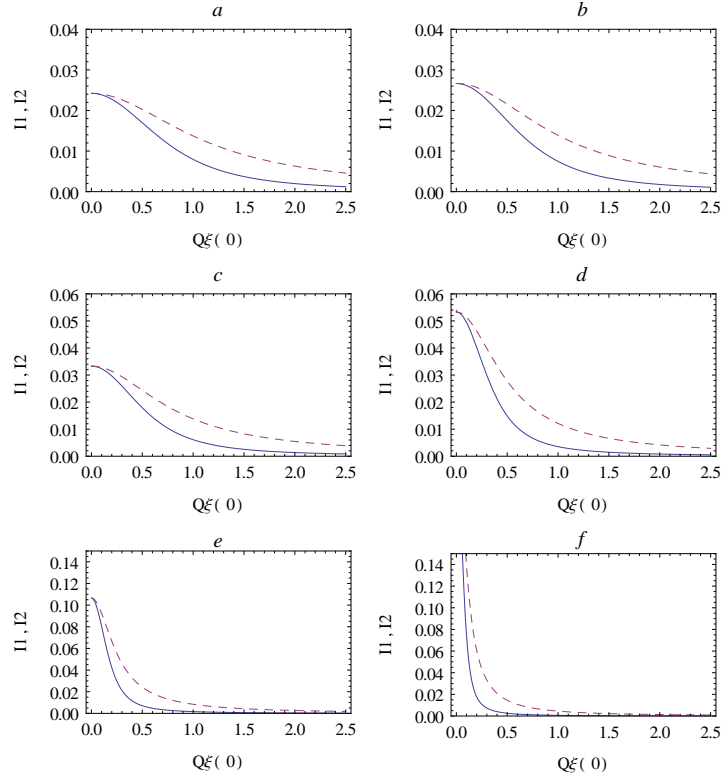}
\end{center}
\caption{(color online) Variations of the classical nonlocal component (i.e. $m=0$) of $I_1$ (full curves) and $I_2$ (dashed curves) with $Q\xi_0$, for different values of the reduced temperature $t$ i.e.: $t=1.1$(a), $t=1$(b), $t=0.8$(c), $t=0.5$(d), $t=0.25$(e), $t=0.1$(f).}
\label{figthree}
\end{figure}

\begin{figure}[tb]
\begin{center}
\includegraphics{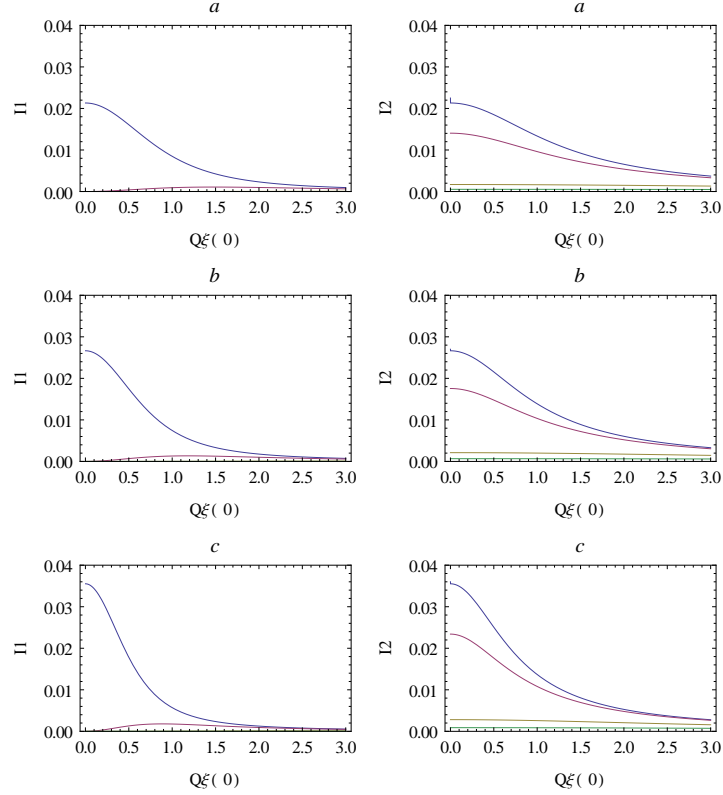}
\end{center}
\caption{(color online) Effects of finite Matsubara frequencies on the variation of $I_1$ (left row) and $I_2$ (right row) with $Q\xi_0$, at different temperature regimes (i.e. $t=1.25$, $1$, $0.75$ for graphs a, b, c respectively). The four curves in each graph correspond to $m=0$, 1, 4 and 8, with the highest maxima of $I_1$ and $I_2$ in each graph corresponding to $m=0$ and the lowest maxima to $m=8$.}
\label{figfour}
\end{figure}

In section~\ref{2} we derived the generalized LGW functional for the BOW instability in the $1D$ electron-phonon system with dispersive phonons, keeping all details from the dynamics of CDW excitations through a space-time dependence of the CDW-CDW correlation function and the mode-mode coupling matrix. A best representation of this dynamic feature is obtained in Matsubara Fourier space where~(\ref{a9}) becomes:
\begin{equation}
\beta\mathcal{F}[\phi]=\sum_{\tilde{Q}}{[\vert g_{ph}^{-1}(\omega_m)\vert + \chi(\tilde{Q},T)]\,\phi(\tilde{Q})\phi^*(\tilde{Q})} + \sum_{\tilde{Q}_1,..., \tilde{Q}_4}{\mathcal{B}(\tilde{Q}_1,...,\tilde{Q}_4, T)\,\phi(\tilde{Q}_1)\phi(\tilde{Q}_2)\phi^*(\tilde{Q}_3)\phi^*(\tilde{Q}_4)}.
\label{a18}
\end{equation}
In this last formula the mode-mode coupling matrix $\mathcal{B}(...)$, which has been obtained explicitely in~(\ref{a13})-(\ref{a15b}), is however here formally represented as a multi-argument function of the fluctuation modes $\tilde{Q}_i$. In fact this multi-argument representation is key for a possible treatment of both momentum and frequency dependent Gaussian fluctuations within the LGW picture, given that we expect the integration of high-energy fluctuations to produce a renormalized low-energy effective field theory for the BO-type CDW instability. The standard procedure in this last respect consists in decoupling the quartic term of the LGW functional to the Gaussian order, with a specific care on momentum-conserving rules and energy scalings among fields $\phi(\tilde{Q})$ coupled via the dynamic mode-mode coupling matrix $\mathcal{B}(\tilde{Q}_1,...,\tilde{Q}_4, T)$. The diagram of figure~\ref{figtwo} represents the proposed fluctuation map in the momentum-frequency space, where fields $\phi(\tilde{Q})$ are assumed to vary with Matsubara frequencies $\omega_m$ on the scale of the characteristic phonon frequency $\Omega$ and in the momentum continuum on the scale $2k_F$. Since every Matsubara frequency $\omega_m$ continuously connects fields along the momentum axis, the lowest pertinent energy scale is the $m=0$ line while high-energy fluctuations, which are to be integrated out, are the sets ($0\leq Q\leq 2k_F, \omega_{m\neq 0}$). \\
With the above separation, a Gaussian decoupling of the quartic term in the LGW functional~(\ref{a18}) leads to the following self-energy: 
\begin{equation}
\Sigma_{HF}(T,\Omega)= \sum_{n=1}^{2}{\Sigma_{HF}^{(n)}(T,\Omega)}, \label{a18a}
\end{equation}
where
\begin{equation}
\Sigma_{HF}^{(n)}(T,\Omega)=\sum_{Q=0}{\sum_{m\neq 0}^{\infty}{\lambda_n\,I_n(\tilde{Q}, T)G_{ren}(\tilde{Q}, T, \Omega)}} \label{a19}
\end{equation}
with $G_{ren}(\tilde{Q}, T, \Omega)$ a phonon-related propagator to be explicitely defined later. Most relevant to underline at this step is the two terms $\Sigma_{HF}^{(1)}(T,\Omega)$ and $\Sigma_{HF}^{(2)}(T,\Omega)$ emerging in the self-energy~(\ref{a18a}), which originate from the possible Gaussian contractions of the two distinct four-point vertices $I_1$ and $I_2$ as explicitely depicted in figure~\ref{figtwoa}. 
We recall that $I_1=I_2$ when ($Q=0, \omega_m=0$) and thus it is in the nonadiabatic and full quantum regime the difference between the two terms in~(\ref{a18a}) is significant. In this regime the Gaussian decouplings of the two distinct four-point vertices  give $\lambda_1=2$ and $\lambda_2=4$. Still these two distinct coefficients are not enough to quantitatively determine the respective weights of $I_1$ and $I_2$ in the self energy~(\ref{a18a}) and in turn in corrections from quantum Gaussian fluctuations. Instead we must resort to a comparision of their variations with the fluctuation momentum and frequency, at different temperature regimes relative to the bare MF critical point $T_c^0$.  \\
On figure~\ref{figthree} and figure~\ref{figfour} we plot $I_1(\tilde{Q}, T)$ and $I_2(\tilde{Q}, T)$ (in units of $LT_c^0/N_F$) versus $Q\xi_0$ for $\omega_m=0$ and different values of the reduced temperature $t=T/T_c^0$ (figure~\ref{figthree}), and at different temperature regimes for diffierent values of the phonon Matsubara frequency $\omega_m$ (figure~\ref{figfour}). The six graphs composing figure~\ref{figthree} suggest that $I_1$ and $I_2$ are equal in the local limit irrespective of temperature as expected. However, when $Q$ is increased $I_1$ decreases faster than $I_2$ above $T_c^0$. Quite remarkably, even at $T=T_c^0$ the two contributions are quantitatively distinct in the presence of finite momenum with $I_2$ always dominant. Far below $T_c^0$ where the neglected quantum dynamics of CDW fields would stabilize a zero-temperature critical point, $I_2$ is the most pertinent contribution to the mode-mode coupling matrix and hence rules the corrections to the CDW transition from Gaussian fluctuations of purely classical nature. \\ 
Concerning their behaviours with simultaneous variation of momentum and frequency at arbitrary temperatures, graphs of figure~\ref{figfour} show that while the magnitudes of $I_1$ and $I_2$ for fixed momentum are both strongly lowered with increasing Matsubara frequency, far below $T_c^0$ all finite-frequency values of $I_1$ are zero and only $I_2$ remains finite hence imposing the nonlocal character of the mode-mode coupling matrix at nonzero Matsubara frequencies. However, numerical calculations show that beyond $m=8$ both $I_1$ and $I_2$ are no more significant, vanishing asymptotically as $t\rightarrow 0$ for all fluctuation momenta. \\   
The behaviours of the mode-mode coupling matrix emerging from figs.~\ref{figthree} and~\ref{figfour} with variations of the fluctuation momentum and Matsubara frequency, suggest two relevant insights about their dynamic properties at different temperature regimes: firstly, although the variation of the mode-mode coupling matrix with $Q$ is quite pronounced in the low-momentum regime it remains significant over the entire $2k_F$ momentum scale when $T$ is close but not equal to $T_c^0$. Secondly, the two contributions behave quite differently with simultaneous variations of the momentum and Matsubara frequency at a given temperature hence their weights in the Hartree-Fock correction should be different. In particular $I_2$ emerges as providing the most significant correction for a quantum-critical point at high Matsubara frequencies, since the contribution from $I_1$ in this frequency regime is relatively smaller over the entire $2k_F$ momentum scale.\\
Having put in evidence the salient features of the two terms contributing to the Hartree-Fock self energy in the presence of dynamic fluctuations, we can now return to formula~(\ref{a19}) and complete the description of quantitites appearing in the dynamic self-consistent equation. \\
The quantity $G_{ren}^{-1}(\tilde{Q}, T, d)$ is defined as:
\begin{equation}
G_{ren}^{-1}(\tilde{Q}, T, d)=m^2(t/\tilde{g}d)^2 + \tilde{a}_r(T, d) + P(\tilde{Q},T), \label{a19b} 
\end{equation}
where
\begin{equation}
P(\tilde{Q},T)= \tilde{\chi}(\tilde{Q}, T) - \tilde{\chi}_{2k_F}(T), \hskip 0.1truecm \tilde{\chi}_{2k_F}(T)= \tilde{\chi}(0, T), \label{a20}
\end{equation} 
$\tilde{\chi}(...)=N_F^{-1}\chi(...)$ and the quadratic coefficient $a_r(T,d)$ of the renormalized GL free-energy is determined via the self-consistent relation:
\begin{equation}
\tilde{a}_r(T, d)=\tilde{a}(T) + \Sigma_{HF}(T,d), \hskip 0.1truecm \tilde{a}(T)=N_F^{-1} a(T). \label{a21}
\end{equation}
Instructively the parameter $d=\Omega/2\pi T_c^0$ is dimensionless, and according to formula~(\ref{a19b}) only for nonzero values of this parameter effects of finite Matsubara frequencies should be sensitive. The formula~(\ref{a19b}) also suggests that quantum effects should be enhanced as $d$ increases at low Matsubara frequencies, a behaviour reflected in the two regimes of the fluctuation map shown in figure~\ref{figtwo}. More explicitely this fluctuation map indicates that quantum effects are significant when the phonon characteristic frequency $\Omega$ is of the order of $2\pi T_c^0$. In fact quantum corrections should be noticeable so long as $\Omega/2\pi T_c^0$ is nonzero, i.e. even at values of this ratio far below $d=1$. 

\section{Quantum corrections}
\label{4}
\subsection{The renormalized critical temperature}
\label{4a}
\begin{figure}[tb]
\begin{center}
\includegraphics{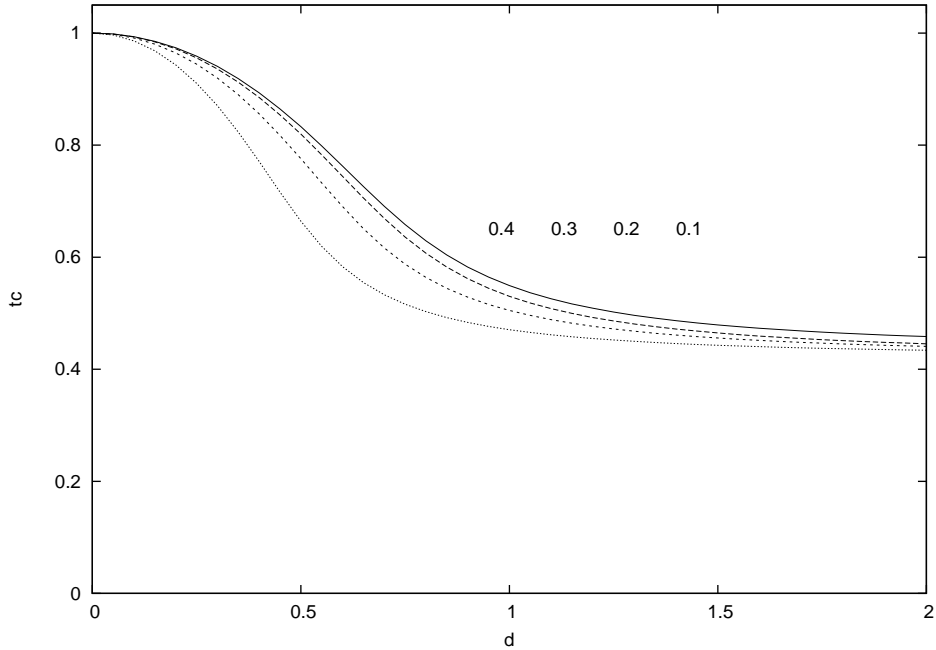}
\end{center}
\caption{$tc$ variation with $\Omega/2\pi T_c^0$ when the two first finite Matsubara frequencies are summed out. Values of $\tilde{g}_0$ are indicated in the graph.}
\label{figfive}
\end{figure}

\begin{figure}[tb]
\begin{center}
\includegraphics{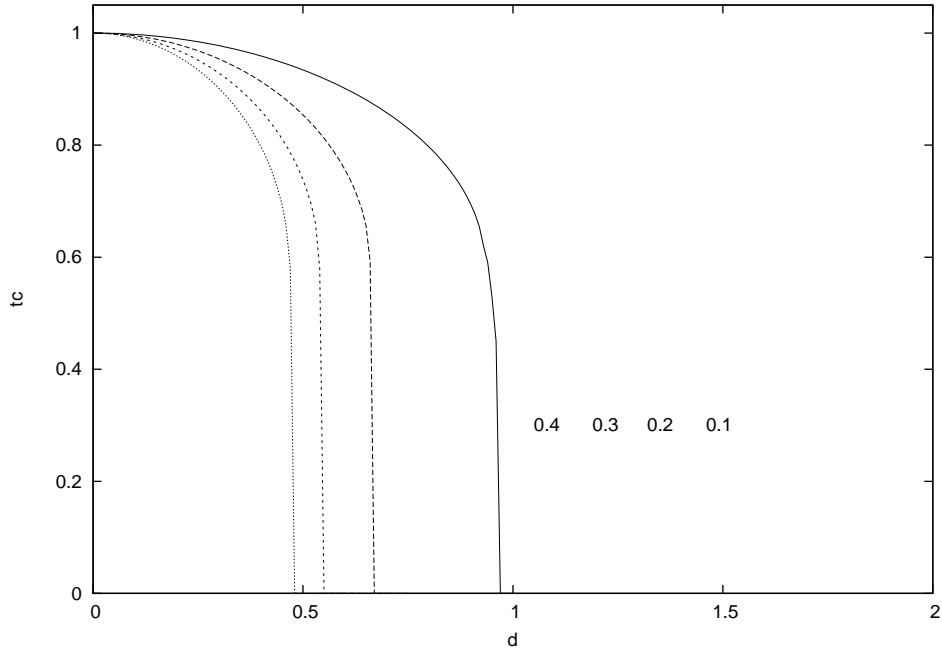}
\end{center}
\caption{$tc$ variation with $\Omega/2\pi T_c^0$ when the eight first finite Matsubara frequencies are summed out. Values of $\tilde{g}_0$ are indicated in the graph.}
\label{figsix}
\end{figure}

We solved numerically the dynamic self-consistent equation~(\ref{a21}) and extracted the renormalized critical temperature $T_c^d$ as a function of the adiabaticity parameter $d$. Although we need not summing all dynamic modes on the fluctuation map (figure~\ref{figtwo}), numerical calculations show that a minimum number of Matsubara frequencies is required for full saturation of quantum corrections. Nevertheless if we restrict the sum to very low frequency fluctuations we achieve only a partial suppression of the transition as figure~\ref{figfive} illustrates, where the reduced temperature $t_c=T_c/T_c^0$ considering only the two first dynamic modes (i.e. $m=1, 2$) is plotted as a function of the adiabaticity parameter $d$ for different values of the effective electron-phonon coupling coefficient $\tilde{g}_0$. 
As we increase the number of Matsubara frequencies we notice more and more strong suppression of the finite-temperature transition, however it is only with the account of $\omega_4$ a finite crossover to the full quantum regime occurs. This finite crossover, however, sharpens with increasing number of Matsubara frequencies and quantum corrections saturate fully in dynamic fluctuations as we sum the $\omega_8$ quantum mode. \\
On figure~\ref{figsix} we represented curves of the reduced renormalized critical temperature $t_c=T_c/T_c^0$ for $\omega_m$ summed up to $\omega_8$, keeping the same values of $\tilde{g}_0$ used in figure~\ref{figfive}. 
We take this opportunity to remark that values of the electron-phonon coupling constant used in our simulations were chosen arbitrarily as our main objective in this study is much more a qualitative analysis. However these values must be sufficiently small for the Fermi energy $E_F$ to remain largely dominant over the characteristic phonon energy and the critical thermal fluctuation $2\pi T_c^0$. This requirement, srictly speaking, is dictated by weak-coupling considerations and has the relevant implication of small values of the ratio $\Omega/2\pi T_c^0$ at the finite crossovers to the quantum critical point. Indeed as curves of figure~\ref{figsix} indicate, in general the ratio $\Omega/2\pi T_c^0$  at the finite crossover to a quantum phase transition is expected to be within the range $0\leq d \approx 1$ in the weak-coupling regime.

\subsection{Thermodynamics in the vicinity of the renormalized CDW transition}
\label{4b}
\begin{figure}[tb]
\begin{center}
\includegraphics{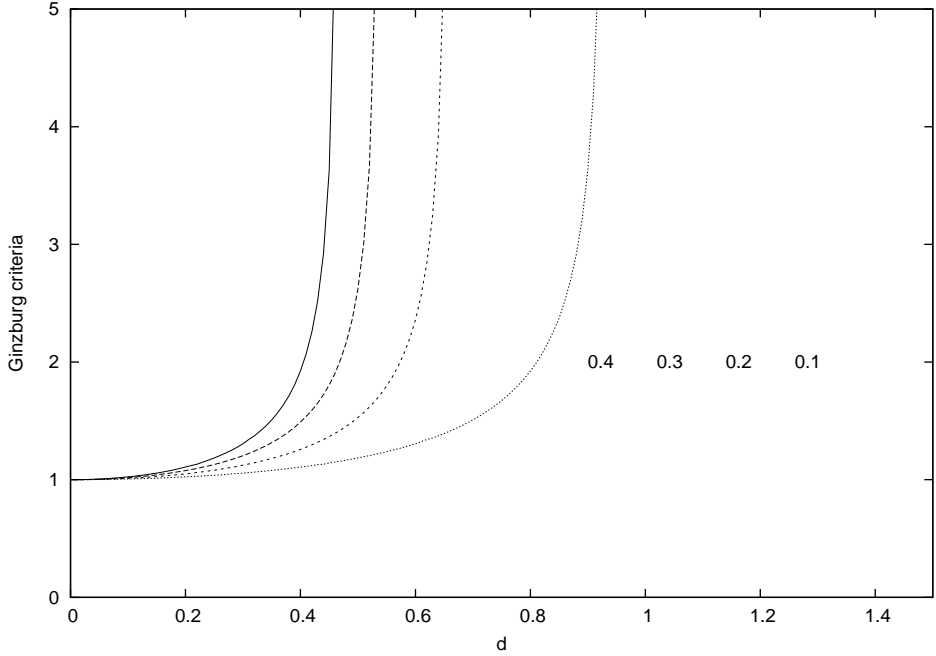}
\end{center}
\caption{Variation of the reduced renormalized Ginzburg criterion with $\Omega/2\pi T_c^0$, for four different values of the effective electron-phonon coupling coefficient $\tilde{g}_0$ indicated in the graph.}
\label{figseven}
\end{figure}

Sufficiently close to the critical temperature, long-wavelength fluctuations become irrelavant and the thermodynamics of the CDW system is dominated by the homogeneous part of the GL free energy. For the renormalized effective field this implies the following free-energy functional:
\begin{equation}
\mathcal{F}_d[\varphi]=\sum_Q{\left[a_r(T,d)\,\vert \varphi(Q)\vert^2 + \frac{b_r(T,d)}{L}\,\vert \varphi(Q)\vert^4\right]}, \label{a22}
\end{equation}
where the microscopic parameters $a_r$ and $b_r$ are of the same forms as~(\ref{a16b}) and~(\ref{a16d}) respectively, except the critical temperature which has been renormalized by dynamic fluctuations and becomes $T_c^d$. Also, because the dynamic self-consistent equation~(\ref{a21}) leads to a nonlinear temperature dependence of the quadratic coefficient $a_d (T,d)$, a linearization of this coefficient around the renormalized critical temperature yields:
\begin{equation}
a_r(T,d)= a_d (\frac{T}{T_c^d}-1), \hskip 0.2truecm a_d= \frac{\partial a_r}{\partial T}\vert_{T=T_c^d}. \label{a23}
\end{equation}
From formula~(\ref{a22}) and~(\ref{a23}) we derive the specific heat jump at the renormalized critical point, which can be expressed~\cite{scalap} as a function of the coherence length $\xi_d$ and the Ginzburg criterion $\Delta t_d$ associate with the new transition point i.e.:
\begin{equation}
\Delta C_d= \frac{1}{\xi_d (\Delta t_d)^{3/2}}. \label{a24}
\end{equation}
$\xi_d$ is inversely proportional to $T_c^d$ and therefore tends to infinity at the crossover value of the adiabaticity parameter $d$. As for the renormalized Ginzburg criterion $\Delta t_d$, a simple calculation shows that this physical parameter can be expressed in terms of its bare MF counterpart as:  
\begin{equation}
\Delta t_d= \frac{\Delta t_0}{a_d}. \label{a25}
\end{equation}
In figure~\ref{figseven}, plots of the reduced quantity $\Delta t_d/\Delta t_0$ indicate a monotoneous increase as the adiabaticity parameter $d$ is increased for the four different values of $\tilde{g}_0$. 
It is particularly remarkable that the renormalized Ginzburg criterion grows infinitely as we approach the quantum critical point, suggesting a very good refinement of the MF theory which turns to be valid over an infinitely wide region around the quantum critical point. On the other hand, according to formula~(\ref{a24}) and~(\ref{a25}) together with the variation of $\xi_d$ with $T_c^d$, the specific heat jump should decrease as the ratio $\Omega/2\pi T_c^0$ is increased in the interval $0\leq d \approx 1$ in the weak-coupling regime, and is expected to vanish asymptotically at the crossover values $d_c$ of the adiabaticity parameter.     

\section{Discussion and concluding remarks}
\label{5}
Investigations on a specific class of materials~\cite{toomb,jerome} displaying peculiar crystal structures, such as organic solids and mixed-valence-band transition-metal compounds, have revealed highly anisotropic electrical conductivities which made them to be considered as $1D$ conductors. These materials are metallic at high temperatures and undergo a metal-insulator transition when the temperature dicreases. A mechanism frequently evoked for this transition is the Peierls instability~\cite{peierls,frohlich} and for materials with a half-filled conduction band, the Peierls instability leads to a linear superlattice distortion producing a CDW excitation and a gap at the Fermi level of the electronic band. Actually the above materials are more pecisely $Q1D$ metals in the sense that they are stacked planes of molecules, with in-plane intermolecular interactions far more stronger than the interaction between planes relatively far apart~\cite{toomb,jerome}. Because of this structure transport properties are dominantly $1D$, and as far as the Peierls mechanism is concerned a finite temperature transition is a manifestation of inter-plane interactions (i.e. $3D$ effects). \\
If the MF theory emerged from past studies as a reasonable framework for a qualitative description~\cite{toomb,jerome,rice,schulz1,mckenzie,leung,rice1} of the Peierls transition in $Q1D$ materials, the theory suffers from the weakness of a long-range CDW order predicted at finite transition temperatures and driven by short-range in-plane intermolecular interactions. Intersting enough, there have been several attempts~\cite{jerome,schulz1} to correct this inconsistency as well as the associate pseudo-gap manifestation in electronic properties~\cite{rice,dumoulin,dumoulina}, by taking account of the contribution of lattice fluctuations when deriving MF characteristic parameters such as the zero-temperature gap. The present work clearly shows that as far as MF considerations are concerned, a coherent and simultaneous treatment of spatial and dynamic features of the lattice fluctuations is necessary in view of a refined effective field theory for the CDW instability, valid irrespective of the temperature range but in the weak electron-phonon coupling regime.       
       
\begin{acknowledgment}
This work is part of the Ph.D. thesis of the first author (University of Sherbrooke, Canada). A. M. Dikand\'e is supported by the Alexander von Humboldt Stiftung, he wishes to thank the Max Planck Institute for the Physics of Complex Systems (Dresden, Germany) where part of numerical simulations was done. The authors are grateful to the Physical Society of Japan for financial support in publication.
\end{acknowledgment}


\begin{thebibliography}{9}
 \bibitem{toomb}G. A. Toombs: Phys. Rep. \textbf{40} (1978) 181.
\bibitem{jerome}D. J\'erome and H. J. Schulz: Adv. Phys. \textbf{31} (1982) 199.
\bibitem{gruner}G. Gr\"uner: \textbf{Density Waves in Solids} (Addison Wesley, 1994)
\bibitem{monceau}P. Monceau: \textbf{Electronic Properties of Quasi-one-dimensional Conductors, I and II} (Dordretch: Reidel, 1995)
\bibitem{peierls}R. E. Peierls: \emph{Quantum Theory of Solids} (Oxford University Press, London, 1955)
\bibitem{frohlich}H. Fr\"ohlich: Proc. R. Soc. London Ser. \textbf{A223} (1954) 296.
\bibitem{rice}P. A. Lee, T. M. Rice and P. W. Anderson: Phys. Rev. Lett. \textbf{31} (1973) 462.
\bibitem{schulz1}H. J. Schulz: in \textbf{Low-Dimensional Conductors and Superconductors}, edited by D. J\'erome and L. G. Caron (Plenum, New York, 1986) p. 95
\bibitem{suh}W. Su, J. R. Schrieffer and A. J. Heeger: Phys. Rev. \textbf{B22} (1980) 2099.
\bibitem{holstein}T. Holstein: Ann. Phys (Leipzig) \textbf{8} (1959) 325. 
\bibitem{dikad}A. M. Dikand\'e: J. Phys. Soc. Jpn. \textbf{78} (2009) 094007.
\bibitem{eliash}A. B. Migdal: Zh. Eksp. Teor. Fiz. \textbf{34} (1958) 1438[Sov. Phys. JETP \textbf{7} (1958) 996]; G. M. Eliashberg: {\it ibid} \textbf{38} (1960) 966[{\it ibid} \textbf{11} (1960) 696.]
\bibitem{mcmillan}W. L. Mc.Millan: Phys. Rev. \textbf{167} (1968) 331.
\bibitem{mckenzie}R. H. McKenzie: Phys. Rev. \textbf{B21} (1995) 6249.
\bibitem{dietrich} W. Dieterich: Adv. Phys. \textbf{25} (1976) 615.
\bibitem{zheng1}H. Zheng, D. Feinberg and M. Avignon: Phys. Rev. \textbf{B39} (1989) 9405.
\bibitem{zheng2}H. Zheng, M. Avignon, and K. H. Bennemann: Phys. Rev. \textbf{B49} (1994) 9763. 
\bibitem{gupta1}P. Sengupta, A. W. Sandvik and D. K. Campbell: Phys. Rev. \textbf{B65} (2002) 155113.
\bibitem{gupta2}P. Sengupta, A. W. Sandvik and D. K. Campbell: Phys. Rev. \textbf{B67} (2003) 245103.
\bibitem{hirsch1}J. E. Fradkin and J. E. Hirsch: Phys. Rev. \textbf{B27} (1983) 1680.
\bibitem{bourbon1}L. G. Caron and C. Bourbonnais: Phys. Rev. \textbf{B29} (1984) 4230.
\bibitem{blawid1}S. Blawid and A. J. Millis: Phys. Rev. \textbf{B63} (2001) 115114.
\bibitem{tam}K. M. Tam, S. W. Tsai, D. K. Campbell and A. H. Castro-Neto: Phys. Rev. \textbf{B75} (2007) 161103.
\bibitem{bakrim}H. Bakrim and C. Bourbonnais: Phys. Rev. \textbf{B76} (2007) 195115.
\bibitem{mermin}N. D. Mermin and H. Wagner: Phys. Rev. Lett. \textbf{17} (1966) 1133.
\bibitem{kotliar}A. Georges, G. Kotliar, W. Krauth and M. J. Rozenberg: Rev. Mod. Phys. \textbf{68} (1996) 13.
\bibitem{hertz}J. A. Hertz: Phys. Rev. \textbf{B14} (1976) 1165.
\bibitem{dumoulin}C. Bourbonnais and B. Dumoulin: J. Phys. I France \textbf{6} (1996) 1727.
\bibitem{nakamura}M. Nakamura: J. Phys. Soc. Jpn. \textbf{68} (2000) 3123.
\bibitem{bourbon2}C. Bourbonnais and L. G. Caron: J. Phys. France \textbf{50} (1989) 2751.
\bibitem{hand1} M. Abramowitz and I. A. Stegun: \textbf{Handbook of Mathematical Functions} (Dover, New York, 1965)
\bibitem{leung}M. C. Leung: Phys. Rev. \textbf{B11} (1975) 4272.
\bibitem{dikathesis} A. M. Dikand\'e: \textbf{Quantum Fluctuations and Structural Instabilities in Low-Dimensional Organic Conductors}, PhD dissertation thesis, University of Sherbrooke, Canada (2004):\\ http://www.usherbrooke.ca/physique/recherche/publications/theses/2004.
\bibitem{scalap}D. J. Scalapino, M. Sears and R. A. Ferrell: Phys. Rev. \textbf{B6} (1972) 3409.
\bibitem{rice1}M. J. Rice and S. Str\"assler: Solid State Commun. \textbf{13} (1973) 125.
\bibitem{dumoulina}B. Dumoulin, C. Bourbonnais, S. Ravy, J. P. Pouget and C. Coulon: Phys. Rev. Lett. \textbf{76} (1996) 1360.
\end{thebibliography}
\end{document}